# One Exploit to Rule them All?
# On the Security of Drop-in Replacement and Counterfeit Microcontrollers

Johannes Obermaier
mail@obermaier-johannes.de

Marc Schink
mail@marcschink.de

Kosma Moczek
kosma@kosma.pl

**Abstract**

With the increasing complexity of embedded systems, the firmware has become a valuable asset. At the same time, pressure for cost reductions in hardware is imminent. These two aspects are united at the heart of the system, i.e., the microcontroller. It runs and protects its firmware, but simultaneously has to prevail against cheaper alternatives. For the very popular STM32F1 microcontroller series, this has caused the emergence of many competitors in the last few years who offer drop-in replacements or even sell counterfeit devices at a fraction of the original price. Thus, the question emerges whether the replacements are silicon-level clones and, if not, whether they provide better, equal, or less security. In this paper, we analyze a total of six devices by four manufacturers, including the original device, in depth. Via a low-level analysis, we identify all of them as being individually developed devices. We further put the focus on debug and hardware security, discovering several novel vulnerabilities in all devices, causing the exposure of the entire firmware. All of the presented vulnerabilities, including invasive ones, are on a Do it Yourself (DiY) level without the demand for a sophisticated lab – thereby underlining the urgency for hardware fixes. To facilitate further research, reproduction, and testing of other devices, we provide a comprehensive description of all vulnerabilities in this paper and code for proofs-of-concepts online.

## 1 Introduction

Well-established markets attract competitors who try to gain a foothold. Products of equal or more benefit for the customer are pushed into the market as drop-in replacements. They usually offer more functionality, a lower price, or even both. Competitors, who join later, often require less investments since the market is already established and they save resources during development by learning from the original product. In its most extreme forms, this corresponds to a precise copying of products, often called product piracy.

Counterfeit and pirated goods reach market volumes of hundreds of billions USD each year with a large share being electronics [14]. On a high-level view, microcontrollers play two major roles for two different stakeholders in terms of product piracy and drop-in replacements: First, there are the users of the microcontrollers who employ them in their embedded systems. The users expect reliable functionality and secure storage of their valuable firmware against unauthorized readout, e.g., by competitors trying to clone the product. Secondly, there are the manufacturers of the microcontrollers who are affected by replacement or counterfeit microcontrollers that could push them, at least partially, out of the market.

The topic becomes especially crucial when the two aspects, replacement microcontrollers and system security, are combined. Often, the source of replacement devices remains unclear: They could either be identical clones, manufactured by somehow obtained manufacturing data, silicon-design reverse-engineering [15], or manually and as-good-as-possible recreated according to publicly available information such as datasheets. Especially in the latter case, key security aspects might have been implemented differently, resulting in better, similar, or worse firmware readout protection.

### 1.1 Related Work

One popular series is the STM32F1 family of microcontrollers by STMicroelectronics (ST). For several years, numerous drop-in replacements, alleging full compatibility, have been appearing, but in different flavors [1, 6, 9–11]. Counterfeit chips, incorrectly marked as legitimate STM32F1, have been reported [7] as well as devices stating their correct origin and claiming only compatibility.

As of 2020, it seems that those companies, mostly founded in the Asian region, try to push into new markets rather offensively. For instance, in an electronics exhibition in Germany, a representative of GigaDevice (GD) approached and outrightly asked the authors of this paper, whether they are using any STM32 products that they could replace. However, those manufacturers commonly disclose very little information on whether the silicon-design is indeed identical and if the security features are sufficiently implemented and tested.

Thus, two questions arise: First, whether the devices are indeed identical to the original microcontrollers or if they are just re-implementations of them. Secondly, and which is even more important, whether these devices provide better, identical, or worse security, compared to the original device.

For the analyzed STM32F1 series, the answer could be unsettling: Although not affected by the STM32F0 vulnerability [13], researchers discovered another hardware design flaw that exposes up to 94 % of the firmware [17]. On the one hand, replacement manufacturers could have (inadvertently) closed this vulnerability, resulting in higher security than in the original device. On the other hand, inapt implementations might cause even bigger issues, decreasing security again.

In this paper, we focus on the security of those devices in depth and shed some light on hardware design authenticity. We acquired five STM32F1 replacement devices of three manufacturers and analyze them together with the original device. We show that each device is a re-implementation of the STM32F1. We discover that all the analyzed replacements, as well as the original, have their *individual* vulnerabilities allowing unauthorized extraction of 100 % of their firmware with low-cost DiY methods and no need for a sophisticated lab.

### 1.2 Contributions

In this paper, we present an in-depth analysis of the original and replacement STM32F1 series comprising:

- A physical and low-level hardware analysis, proving that each device is an independent re-implementation
- An in-depth security analysis of the original device and five replacements
- Discovery and description of more than ten vulnerabilities, allowing for 100 % firmware extraction from every device, i.e.,
  - Multiple severe debug interface vulnerabilities
  - Invasive hardware attacks on multi-die systems
  - A power glitch exploiting software live-patching
- A demonstration that the attacks, including the invasive ones, can be brought down to a low-cost DiY level
- Proofs-of-Concepts for each discovered vulnerability, also published online together with further materials

### 1.3 Structure

In the next section, we state the scope of analyzed devices. We continue with an overview of the security concept in Section 3. First, we take a detailed look at the physical differences of the microcontrollers in Section 4. Next, in Section 5, we derive software and hardware attack vectors on the devices with the focus on DiY approaches. Then, we analyze all devices for debug interface vulnerabilities and describe their exploitation in Section 6. Additionally, we demonstrate invasive attacks in Section 7. We discuss countermeasures in Section 8 and conclude our paper in Section 9.

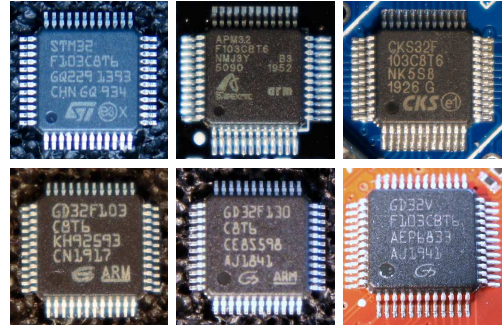

Figure 1: The Devices under Test, listed in Table 1.

## 2 Devices under Test

For our analysis, we chose five replacements and an original device of the very popular STM32F103, a member of the STM32F1 family. They are shown in Figure 1 and listed in Table 1. In 2020, all these devices were available on the free market, sometimes advertised as STM32F103 drop-in replacements. The devices are based on an ARM Cortex-M3, except for the GD32VF103 that instead contains a RISC-V core. The GD32F130 is a special case as it does not have a direct STM32 counterpart; however, it appears to be a refined GD32F103 with additional capabilities, and most interestingly, extended security features. Thus, we analyze a broad range of devices form drop-in replacements to upgrades.

The extended chip names, given in braces in Table 1, primarily denote the available memory and IC package. We experimented mainly with *C8T6* devices containing 64 KiB flash memory. If unavailable, we switched to *CBT6* devices offering twice the flash memory. Although not tested due to the sheer number of device variations, our results will very likely apply to other variants as well.

While acquiring our devices under test (DuTs), we noticed that some vendors undercut the usual price of an STM32F103 by far; thus, we ordered a batch for analysis. Although components marked with STM32F103C8T6 were delivered, the debugger warned us that the IDCODE 0x2BA01477 does not match the expected value of 0x1BA01477. Under strong illumination and by tilting the device, a very faint *CKS* marking appears, as shown in Figure 2. This part of the marking exactly matches an actual CKS32F103 device, hence, counterfeit STM32F103 with a manipulated marking were delivered. Thus, users have a risk of unintentionally employing counterfeit devices in their systems – including their vulnerabilities that we describe in the paper.

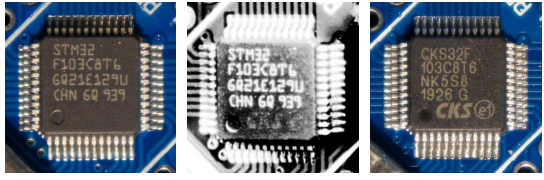

Figure 2: Under special illumination and image enhancements (center), the counterfeit STM32F103 (left) reveals the *CKS* marking of the actual chip (right).

Table 1: List of devices selected for analysis.

| Device | Manufacturer |
|---|---|
| STM32F103(C8T6) | STMicroelectronics |
| APM32F103(CBT6) | Apex Microelectronics |
| CKS32F103(C8T6) | China Key System & Integrated Circuit |
| GD32F103(C8T6) | Gigadevice |
| GD32F130(C8T6) | |
| GD32VF103(CBT6) | |

## 3  Security Concept

The confidentiality of the internal flash memory is the main aspect in the microcontrollers' security concept. For that, the RDP level represents the security configuration of the device, defining the debug interface permissions [19]. All analyzed devices support two RDP levels, except the GD32F130 which has three levels [11]. In level 0, the chip is entirely unprotected and flash memory can be read and written by the debugger. This mode is used during development, for example. In level 1, the debug interface is still enabled and SRAM is accessible, however, debug access to flash memory will be blocked and the firmware remains protected. In level 2, only supported by the GD32F130, the debug interface is shut down, thus, neither a debugger can be attached, nor flash memory is readable.

The RDP configuration is stored as a 16-bit value in the option bytes region within the flash memory. Its value is loaded once during device startup. For level 0, the value is 0xA55A, and for level 2, the value is 0xCC33. All other values map to level 1. Upgrading security is always possible, downgrading security triggers firmware erasure in level 1 and is impossible in level 2.

The device supports booting from SRAM and flash memory, selected by the *BOOT0* pin. Usually, the signal is set to 0 and the device boots from flash memory. If driven to 1, the device boots a temporary firmware previously loaded into the SRAM. If at least RDP level 1 is set and SRAM-boot is selected, access to flash memory is prevented [19].

## 4  Hardware Analysis

In this section, we identify similarities and differences between the devices, discovering verbatim copied parts and individual developments.

### 4.1  Integrated Bootloader

Each device contains a bootloader firmware that is pre-programmed by the manufacturer. A byte-wise comparison of the binaries shows that the bootloader firmware of the APM32F103, the CKS32F103, and the STM32F103 are more than 97% identical. While having identical instructions, they only differ in a few data constants such as serial numbers. Thus, the bootloader has clearly been copied from the STM32F103 to the others.

In contrast to this, the bootloaders of the GD32F103 and GD32F130 do not show any significant similarities to other devices. Although the bootloader implements a similar functionality, it appears to be an own implementation and has not been copied. The same is true for the bootloader of the GD32VF103, since it is the only RISC-V based device. Altogether, verbatim copying of the bootloader has been observed for some, but not all devices.

### 4.2  CPU Core Revision

The Cortex-M3 core supports a CPUID register that indicates its hardware revision and patch number. The STM32F103 shows a value of 0x41**1**FC231, corresponding to revision 1. In contrast to this, the APM32F103, CKS32F103, GD32F103, and GD32F130 all indicate 0x41**2**FC231, i.e., they are based on the more recent CPU core (revision 2).

We verified that the indicated CPU revisions are correct and not feigned. We successfully verified that the CPU hardware erratum "532314: DWT CPI counter increments during sleep" [4] is not present in the devices claiming revision 2. Additionally, the size of the CPU's vector table offset register (VTOR) in these devices indeed supports two further bits [3], compared to the previous revision [2]. Thus, the replacement devices contain a more recent CPU core revision compared to the STM32F103, strongly hinting at new developments. However, this does still not allow to conclude whether *all* of them are individual developments or just distributed under different names.

### 4.3  Decapping of Devices

By opening the packages of the microcontrollers, called decapping, the internal silicon dies are exposed and become visible for optical inspection to identify identical devices.

For the experiment, we heated the chips up with a laboratory hot plate and repeatedly applied sulfuric acid. After some time, we were able to scrape away the top of the casing and

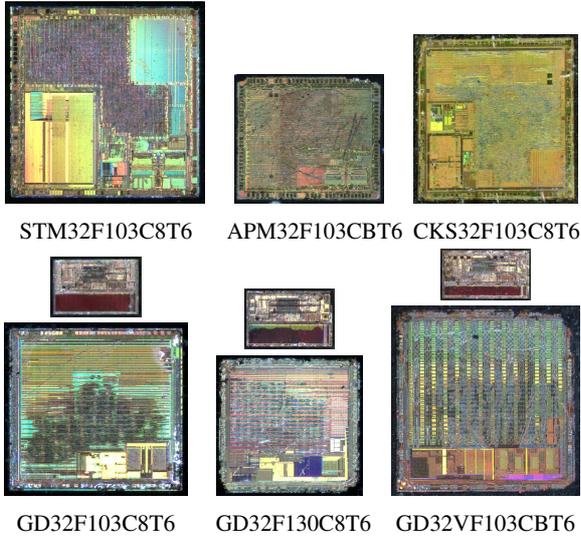

STM32F103C8T6  APM32F103CBT6  CKS32F103C8T6

GD32F103C8T6  GD32F130C8T6  GD32VF103CBT6

Figure 3: The silicon dies after decapping, true to scale; the exact dimensions are listed in Table 3 in Appendix A.

exposed the die. The chips are not functional anymore, but the dies stay intact as they are covered by a passivation layer.

The chips, shown in Figure 3, are very different from each other, thus, all of them are individual developments. The sizes of the dies vary widely, as listed in Table 3 in Appendix A, even hinting at different manufacturing technologies. The GD32 devices contain two dies in their package which was seldom observed in microcontrollers. Further analysis showed, in accordance with other analyses [12, 21], that the smaller chip contains the flash memory and the larger one incorporates the remaining logic such as the CPU, SRAM, and peripherals.

## 5 Analysis Approach and Attack Vectors

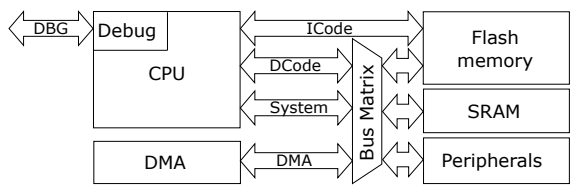

Figure 4: Simplified block diagram of the internal buses of a typical microcontroller.

The confidentiality and integrity of the flash memory contents, i.e., the user's data and firmware, is the primary goal of the security concept. Since the previous physical comparison showed no major similarities, the devices are analyzed for security individually. At first, we analyze the attack vectors via the debug interface for devices set to RDP level 1. In the system, a multitude of components is connected to the internal buses, as shown in Figure 4. The debug module, the CPU, and the Direct Memory Access (DMA) module are masters on the buses, thus, they can perform read-transactions from flash memory. For that, we test whether in RDP level 1,

- **D0**: the *debugger* can directly read out the memory. This shows if security measures are implemented at all.
- **D1**: the *CPU* can be forced to leak flash data via executing load instructions or interrupt vector fetches.
- **D2**: the *DMA* can be forced to read from flash memory.

Regarding hardware-based attacks, we focus on low-cost attacks which are executed with basic, easily available, and comparably cheap tools. For that we test whether,

- **H0**: the debug interface is indeed shut down in RDP level 2 (GD32F130 only).
- **H1**: data can physically be *eavesdropped*.
- **H2**: data or RDP levels can physically be *manipulated*.
- **H3**: power glitches can facilitate firmware extraction.

## 6 Debug Interface Analysis, Vulnerabilities, and Exploits

In this section, we analyze the security implementations, outline the vulnerabilities, and demonstrate their exploitation. All vulnerabilities, grouped by devices, and their CVEs are summarized in Table 4 in Appendix D.

### 6.1 D0: Direct Debug Access

**Affected:** None

After setting each device to RDP level 1 and performing a power cycle, the flash memory is not readable via a debugger anymore. Thus, all devices implement basic security features that protect the firmware against direct readout.

### 6.2 D1-A: Load Instruction Exploitation

**Affected:** CKS32F103, GD32VF103

Although a debugger is prevented to *directly* read the flash memory, this is not always true for the CPU [5, 16, 17]. Even if a device is in RDP level 1, debuggers are allowed to read and write CPU registers, including halting and starting execution. However, direct data reads via the CPU from flash memory are mostly blocked to prevent firmware extraction.

We discovered that this does not apply to the CKS32F103 and GD32VF103. In both systems, load instructions, e.g., `ldr` for ARM and `lw` for RISC-V, are not sufficiently blocked if a debugger is present.

For the GD32VF103, firmware being executed from flash memory or SRAM is always allowed to read from the flash memory. Thus, for exploitation, an attacker loads an extraction firmware into the SRAM. It comprises an `lw s5, 0(s1)` instruction that loads data from the flash address in register `s1` and copies it into register `s5`. The extraction firmware then transmits the data to the attacker via a UART interface. The extraction firmware repeatedly executes this instruction while sweeping the address `s1` over the entire flash memory, thereby extracting the flash memory contents within a minute.

The CKS32F103 requires an alternative approach as only code in the flash memory but not in the SRAM is allowed to access the flash memory. Thus, the attacker needs to know the location of an `ldr` instruction in the flash memory beforehand – what appears to be a chicken-and-egg problem. Although trying to guess the location of an `ldr` is possible, reusing the integrated bootloader makes the approach trivial. The bootloader is always readable and contains several suited gadget-instructions such as `ldr r1, [r0]`. It loads data from the address in register `r0` and copies it into the register `r1`. By repeatedly executing this instruction via a debugger, while sweeping the address in `r0` over the entire flash memory, the firmware is extracted within ten minutes.

Both approaches were fully automated, including `ldr`-gadget detection and exploitation. The remaining microcontrollers did not exhibit this vulnerability. We tested various further ARM data-access instructions including `pop`, `ldrb`, `ldrh`, `tbb`, and `tbh`, without success.

## 6.3 D1-B: Extraction via Exceptions

**Affected:** STM32F103, APM32F103, CKS32F103

As illustrated in Figure 4, the flash memory on an ARM Cortex-M based device is accessed via the data bus and instruction bus. The data bus serves data and debug accesses; the instruction bus is intended for instruction and interrupt vector fetches from flash memory [2].

We noticed that while a debugger is present in RDP level 1, access to the flash memory is only blocked for the data bus but not for the instruction bus. Because interrupt vector fetches are performed via the instruction bus, they always succeed even with a debugger connected – resulting in a security weakness.

On exception entry, the CPU fetches the corresponding interrupt handler address from the interrupt vector table in flash memory and loads this address into the program counter (PC) register, accessible via the debugger. This allows an adversary to read out flash memory contents in the vector table by deliberately generating exceptions via the debug interface and observing the loaded vector address. By reconfiguring the CPU via the vector table offset register (VTOR) [2], the table is relocated and the vectors are fetched from this alternative address – thereby exposing further flash memory contents.

An example vector table, relocated to different addresses in flash memory, is depicted in Figure 5. The beginning of the vector table is determined by the VTOR. The number of interrupts is implementation defined and varies among the different devices. Some entries in the vector table, marked in red, are unused and therefore inaccessible by the processor.

Figure 5: Vector table moved to another location in flash memory via the VTOR. Inaccessible table entries are highlighted.

For that reason, the attack is limited and the flash memory cannot be extracted entirely. However, by taking advantage of the fact that external interrupts outside of the vector table are *wrapped around* and mapped to the beginning of the vector table, the number of inaccessible table entries is reduced.

This attack was already demonstrated for the STM32F1 series [17]. We tested this approach on further devices and discovered that the APM32F103 and CKS32F103 series are also affected. The extraction performance, listed in Table 2, is comparable. The increasing coverage results from the fact that more interrupts allow extracting inaccessible table entries by exploiting the *wrap-around* behavior. Altogether, the results show that up to 93.8 % of the flash memory can be extracted, taking less than an hour.

Table 2: Duration and coverage of the firmware extraction process via deliberate exception generation.

| Device    | Interrupts | Extraction time | Coverage |
|-----------|------------|-----------------|----------|
| STM32F103 | 59         | 48 min          | 89.1 %   |
| APM32F103 | 75         | 52 min          | 93.8 %   |
| CKS32F103 | 76         | 53 min          | 93.8 %   |

## 6.4 D1-C: VTOR Control Flow Redirection

**Affected:** GD32F103

We noticed that the GD32F103 continues execution even when a debugger is attached. However, when the CPU debug module is enabled (`C_DEBUGEN` bit in the `DHCSR` register [2]), e.g., for halting the CPU or accessing processor registers, flash memory access becomes locked down for all bus masters, including the CPU, and execution stops.

However, even if the `C_DEBUGEN` bit is not set, the remaining *system* level components, such as SRAM and peripherals, are still accessible since they are not part of the CPU. Thus, one is allowed to write a flash memory dumping firmware into an unused region of the SRAM. Nevertheless, the CPU cannot *directly* be set to execute this firmware, as writing to the program counter register via the debugger will trigger the aforementioned mechanism.

Instead, we *indirectly* redirect the control flow via the VTOR. We update the VTOR to point to our flash dumping firmware. While the CPU is executing its original firmware, we trigger a Non-Maskable Interrupt (NMI) which forces the CPU to jump into the NMI handler – which is actually our flash memory dumping firmware in the SRAM. Please note that exceptions are executed in privileged mode, allowing deepest system access. Furthermore, the NMI cannot be interrupted by any other exception nor disabled by software. Thus, the extraction firmware runs unimpeded. Due to this vulnerability, the entire firmware is exposed in less than a minute.

## 6.5 D2: DMA Access Exploitation

**Affected:** CKS32F103, GD32F103

The DMA module is a master on internal buses and designed to quickly move data from one peripheral to another without CPU interaction. In most implementations, the DMA is not allowed to access the flash memory anymore as soon as a debugger is attached. However, we noticed that this is not correctly implemented in the CKS32F103 and the GD32F103.

In the CKS32F103, the DMA module is always allowed to read from flash memory. However, in the GD32F103, DMA access to flash is only possible as long as the CPU debug module has not been activated, similar to Section 6.4.

In both cases, the debugger configures the DMA to copy flash memory contents into a directly readable memory such as SRAM. For that approach, the DMA is set to memory-to-memory mode, with the flash memory as source and the SRAM as destination. Next, the DMA transactions are started. As the SRAM is not sufficiently large to incorporate the entire flash memory, the firmware is extracted in multiple chunks.

For the CKS32F103, the CPU should be halted to prevent any interference of SRAM accesses during the DMA-based copy procedure. However, the CPU must *not* be halted for the GD32F103 as this would trigger the protection mechanisms and lock down the flash memory. Instead, we prevent inter-ference by deliberately crashing the CPU, thereby halting the CPU without triggering the protection mechanism. To do so, we reconfigure the VTOR to point to an invalid memory address, such as `0xF0000000`. Then, we trigger an NMI which leads to a CPU crash as this address can neither be read nor executed from. This vulnerability exposes the entire users firmware within a few minutes.

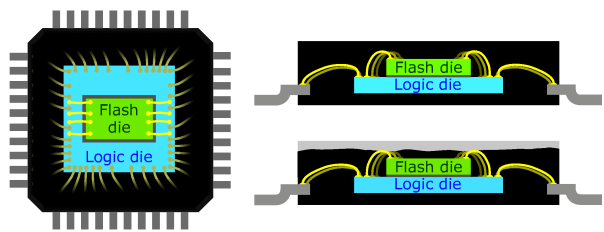

Figure 6: Top and side view of a dual-die device. The bonding wires between both dies become accessible by partially removing the casing (right, bottom).

## 7 Physical Analysis and Exploits

### 7.1 H0: Debug Access in RDP Level 2

**Affected**: None

For the experiment, we set the GD32F130 to RDP level 2, which is only supported by this device. After cycling power, the debugger is unable to connect to the device anymore. Thus, RDP level 2 is implemented in this device.

### 7.2 H1: Invasive Data Eavesdropping

**Affected**: At least GD32F103, GD32F130

All tested GD32 devices have a separate logic and flash memory die, as shown in Section 4.3. Both chips are stacked atop of each other and connected via bonding wires, illustrated in Figure 6. This sparked the question, whether such a design is sufficiently secure against invasive physical attacks – especially in the case of the GD32F130 that did not exhibit any debug interface vulnerabilities so far.

In order to gain access to the bonding wires for eavesdropping, the top of the casing is carefully removed via sand paper of 120 and 600 grit size. A small portion of the bonding wires' top curvature becomes sufficiently accessible to contact it momentarily with a very thin wire. Without requiring a pre-amplifier, these signals are connected to a logic analyzer.

We identified a Quad-SPI (QSPI) bus at 4 MHz between the logic and the flash die. *Quad* means that the bus communicates via a 4-bit wide parallel interface for addresses and data. To record all data, all four signals must be accessed.

To create a larger, electrically improved, and mechanically more stable contact area for further analysis, we applied conductive silver paint. A fine brush, manually made from a sin-gle fiber of a bamboo stick, is helpful to apply the paint. Via additional thin wires, all four QSPI signals are permanently connected to the circuit board. The result of this rather tedious task is shown in Figure 7. Due to the regular communication pattern and availability of a datasheet of a functionally similar standalone QSPI flash device by GD [8], no additional signals need to be probed for reverse-engineering.

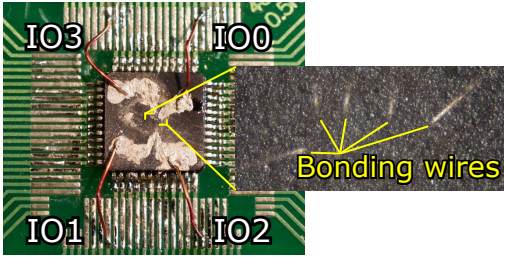

Figure 7: The prepared GD32F130 with access to all four QSPI data signals (IO0 to IO3).

After waking up the flash memory, the *factory configuration*, containing a device ID and most likely some calibration data, is loaded. Next, the integrated *bootloader* section is read. Directly afterwards, the device fetches the *option bytes*, containing the security settings. Then, the *firmware* is fetched in pages of 1 KiB. The GD32F103C8T6 initially fetches the entire memory of 64 KiB and very likely copies it to an internal SRAM for faster code execution [21]. The GD32F130C8T6 fetches only 32 KiB on start up and the rest on demand. While the factory configuration, the bootloader, and the option bytes are clearly visible, the firmware appears scrambled and obfuscated. We assume that this is intended as a security measure and not as a memory optimization.

To decode the firmware, we reverse-engineer the obfuscation mechanisms. For the analysis, we use the following flash address structure of a 64 KiB device:

$$\text{addr} = \underbrace{00001000}_{\text{Main Flash}}\ \underbrace{00000000\ \text{PPPPPP}}_{\text{Page}}\underbrace{\text{AA\ AAAAAA}}_{\text{Word}}\underbrace{\text{BB}}_{\text{Byte}}$$

First, we noticed that the page *address*, defined by bits 10 to 15, is not obfuscated nor scrambled. Pages filled with data appear in the same order on the QSPI bus as they are present in memory, albeit the data inside each page is scrambled.

### 7.2.1 Word-Permutation

However, we observed that some pages, especially the only partially filled ones, show a regular pattern during their QSPI transmission. Several larger chunks of unused memory were interspersed in the firmware. Thus, we suspected and confirmed a page-internal permutation of data words.

Let $\mathbf{a}_P$ be the column vector of bits representing the physical address of a word inside a flash memory page. Similarly, $\mathbf{a}_L$ is the column vector of bits representing the *logical* address of the word in the system's logical view of the page. Since each page is 1 KiB large and word-addressed, there are 8 address bits.

$$\mathbf{a}_P = [a_{P9}\,a_{P8}\,a_{P7}\,a_{P6}\,a_{P5}\,a_{P4}\,a_{P3}\,a_{P2}]^T,\quad \mathbf{a}_P \in \mathbb{Z}_2^{8\times 1} \quad (1)$$

$$\mathbf{a}_L = [a_{L9}\,a_{L8}\,a_{L7}\,a_{L6}\,a_{L5}\,a_{L4}\,a_{L3}\,a_{L2}]^T,\quad \mathbf{a}_L \in \mathbb{Z}_2^{8\times 1} \quad (2)$$

To reverse-engineer the scheme, we place one all-zero word at the beginning of individual pages with an offset of: `0x004`, `0x008`, `0x010`, `0x020`, `0x040`, `0x080`, `0x100`, `0x200`. This corresponds to addresses which have only one address bit set, i.e., $a_{L2}$, $a_{L3}$, $a_{L4}$, $a_{L5}$, $a_{L6}$, $a_{L7}$, $a_{L8}$, and $a_{L9}$.

By observing the resulting address of the all-zero words within the QSPI transactions, the permutation of the words and thereby the address bits are derived. After verifying the permutation using additional addresses in tests, the permutation from the logical order to the physical order can be expressed via the permutation matrix $\mathbf{P}_{\text{aPL}}$ as:

$$\mathbf{a}_P = \mathbf{P}_{\text{aPL}} \cdot \mathbf{a}_L, \quad \text{with } \mathbf{P}_{\text{aPL}} \in \mathbb{Z}_2^{8\times 8}. \quad (3)$$

For the GD32F130C8T6, we achieved the following permutation matrix:

$$\begin{bmatrix} a_{P9} \\ a_{P8} \\ a_{P7} \\ a_{P6} \\ a_{P5} \\ a_{P4} \\ a_{P3} \\ a_{P2} \end{bmatrix} = \begin{bmatrix} 0 & 1 & 0 & 0 & 0 & 0 & 0 & 0 \\ 0 & 0 & 0 & 0 & 0 & 0 & 1 & 0 \\ 0 & 0 & 0 & 0 & 1 & 0 & 0 & 0 \\ 0 & 0 & 1 & 0 & 0 & 0 & 0 & 0 \\ 1 & 0 & 0 & 0 & 0 & 0 & 0 & 0 \\ 0 & 0 & 0 & 0 & 0 & 0 & 0 & 1 \\ 0 & 0 & 0 & 0 & 0 & 1 & 0 & 0 \\ 0 & 0 & 0 & 1 & 0 & 0 & 0 & 0 \end{bmatrix} \cdot \begin{bmatrix} a_{L9} \\ a_{L8} \\ a_{L7} \\ a_{L6} \\ a_{L5} \\ a_{L4} \\ a_{L3} \\ a_{L2} \end{bmatrix} \quad (4)$$

The permutation matrix for the GD32F103C8T6 is given in Appendix B. The inverse, which is the transposed matrix, provides the conversion from physical to logical addresses.

### 7.2.2 Bit-Permutation

After reverting the word-permutation, the data is still not directly readable as the data is again obfuscated. Though, the hamming weight of each word is correct. Additionally, a memory region filled with identical bytes will lead to a repetitive pattern of 32 bits in length. This leads to the conclusion that no strong encryption is present but only a permutation of bits inside a block of 32 bits.

Due to the vast number of possible permutations, the mapping is inferred via a test pattern written into the flash memory. To become independent of the address bit permutation, each test-word is written to the beginning of a new page. The following test pattern, in which each column encodes its bit-index within the word, is written to memory in big endian encoding:

```
0xFFFF0000 (11111111111111110000000000000000)
0xFF00FF00 (11111111000000001111111100000000)
0xF0F0F0F0 (11110000111100001111000011110000)
0xCCCCCCCC (11001100110011001100110011001100)
0xAAAAAAAA (10101010101010101010101010101010)
```

Next, the corresponding five QSPI transactions loading these words from flash memory are recorded and analyzed. When looking at a specific bit in all five transactions, a unique binary pattern is seen. The MSBs of the five transactions, for example, are `11011` in binary which is 27 in decimal. This means that bit 27 of the data is actually being transmitted as the MSB on the bus. By repeating this analysis for the remaining bits, the entire bit-mapping can quickly be inferred. As a sanity-check, the resulting mapping must be bijective.

In all devices analyzed by the authors so far, permutations were only done within the same byte. Thus, this permutation can be split up into four independent 8-bit permutations for simplicity. However, these four permutations are not identical.

Let $\mathbf{d}_P$ be the column vector of the physical bits of a byte on the QSPI bus. Let $\mathbf{d}_L$ be the column vector of the logical bits of a byte inside the system. $\mathbf{P}_{dPL}$ is the permutation matrix to convert the data bits from the logical order to the physical order.

$$\mathbf{d}_P = \mathbf{P}_{dPL} \cdot \mathbf{d}_L, \quad \text{with} \quad \mathbf{d}_P, \mathbf{d}_L \in \mathbb{Z}_2^{8\times 1}, \; \mathbf{P}_{dPL} \in \mathbb{Z}_2^{8\times 8} \quad (5)$$

The four permutation schemes, explicitly provided in Appendix C and identical for the GD32F103 and GD32F130, are used in cyclic order depending on the addresses' LSBs, i.e.,

$$\mathbf{P}_{dPL} = \begin{cases} \mathbf{P}_\alpha, & \text{if addr}[1..0] = \texttt{00} \\ \mathbf{P}_\beta, & \text{if addr}[1..0] = \texttt{01} \\ \mathbf{P}_\gamma, & \text{if addr}[1..0] = \texttt{10} \\ \mathbf{P}_\delta, & \text{if addr}[1..0] = \texttt{11} \end{cases} \quad (6)$$

The inverse of the matrix, which is the transposed matrix, provides the conversion from physical to logical order.

The results show that the obfuscation via permutation is a weak security measure as the matrices can be entirely extracted with little effort. This enables an attacker to extract and decode the firmware, thereby circumventing device security.

### 7.3 H2: Invasive RDP Manipulation

**Affected:** At least GD32F130

The GD32F130 supports RDP level 2 which entirely shuts down the debug interface. Additionally, no vulnerabilities are known for this device in RDP level 1. Thus, an attacker would have to focus on hardware attacks. However, sniffing QSPI communication is a viable but effortful approach due to the difficult probing of bonding wires and the de-obfuscation required for data extraction.

Thus, an alternative for getting access to the firmware is an active manipulation of QSPI communication. Analysis showed that the RDP setting is loaded during startup and present on the bus in plain without obfuscation. The naive approach would be to interfere with the option byte loading by directly manipulating the option byte transmission. Downgrading from RDP level 2 to 1 to re-enable the debug interface is feasible by flipping any bit, however, this does not provide access to the firmware as no debug interface vulnerabilities are known for level 1. To downgrade from level 2 to 0, the option bytes would have to be altered from `0xCC33` to `0xA55A`. While being at least theoretically possible, this would again require access to all four QSPI data signals at once.

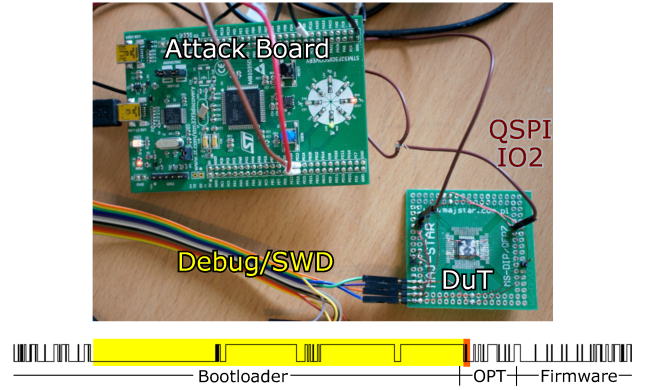

Figure 8: QSPI active manipulation setup with transaction excerpt. A fault (red) is injected into the IO2 signal to alter the option bytes, aligned to the distinct pattern (yellow) at the end of the bootloader.

However, detailed analysis shows that the bus traffic contains the value for RDP level 0, which is `0xA55A`, at the beginning of the *factory configuration* QSPI transaction:

**`A55A`** `07EF 1004 0313 FFFF [....]`

The reason for this is unknown but it enables a new attack vector. Further analysis shows that the factory configuration is loaded from address `0x0400` and the option bytes are loaded from address `0x4000`. Thus, by manipulating the *address* of the QSPI transaction loading the option bytes, they are forced to be loaded from the factory configuration region instead. This replaces the actual option bytes with the value `0xA55A` and the device falls down to RDP level 0 – enabling full access to the entire user firmware via the debugger.

The physical manipulation is comparably simple, as only two bits have to be flipped. Furthermore, both bits to be manipulated are transmitted via the QSPI IO2 signal, thus, an attacker only has to gain access to this *single* bonding wire.

The attack is executed by an STM32F303 discovery board running at 72 MHz. This attack board runs a timing-optimized assembly implementation of this address faulting approach. Due to the high currents required for this forceful override, two GPIOs of the attack board are connected in parallel. The system is connected to the QSPI IO2 signal, actively monitors it, and triggers on the address phase of the option byte loading, as shown in Figure 8. The correct point in time is identified by waiting for the typical pattern of many zeroes followed by three block of ones that occurs only at the end of the bootloader transaction. The address phase of the option byte loading starts exactly in the moment when the IO2 line is actively pulled low again. At this point in time, the attack board enables its output drivers, overrides the address with `0x0400`, and enters high-impedance mode again. Next, the QSPI transaction returns the value `0xA55A` for the option bytes instead of the actual value, thereby falling back to RDP level 0. This re-enables the debug interface without any protection, hence, the firmware is directly readable.

Although the firmware is protected in RDP level 2 or 1, this approach disables security entirely. When correctly set up, the attack works very reliable even on the first try. Despite the forceful override, this attack has not caused any damage to the DuT. Please note, that this is a low-cost invasive attack with easily available materials, thus, the bar for an attacker is very low. This demonstrates the conceptual weakness of multi-die systems that have accessible inter-die bonding wires.

## 7.4 H3: Shellcode Exec. via Glitch and FPB

**Affected:** APM32F103, STM32F103

In this section, we present a novel multi-stage low-level attack that spawns a privileged shell on a secured microcontroller. This allows us to read out the protected flash memory and to do arbitrary modifications to the system. The basic idea is to repurpose a hardware-based firmware live-patching mechanism to redirect the control flow from the original firmware into our shellcode in SRAM. For that, we apply a supply voltage glitch, trigger the execution of a two-stage exploit code in SRAM, and repurpose the core's firmware live-patching mechanism.

### 7.4.1 Background

The Cortex-M3 core comprises the Flash Patch and Breakpoint Unit (FPB) which is a hardware module intended for software live-patching as well as debugging [3]. The module has address comparators that trigger when a specified memory address is accessed by the core. The actual data can then be replaced with alternative contents. Since breakpoints and patches should survive a device reset, this module keeps its configuration under reset and is only cleared by a power cycle.

To determine whether to **boot** from flash memory or SRAM, the BOOT0 and BOOT1 pins are sampled after reset [20]. When SRAM-booting is selected, this firmware must be loaded to SRAM beforehand, as SRAM is a volatile memory and contents become lost at power-off. Please note that when booting from SRAM in RDP level 1, the flash memory is inaccessible.

Independent of the bootmode, if a debugger is attached to the system, flash memory is **locked down**, preventing any further access [19]. The flash memory lockdown is only released again by a power cycle.

### 7.4.2 Attack Execution

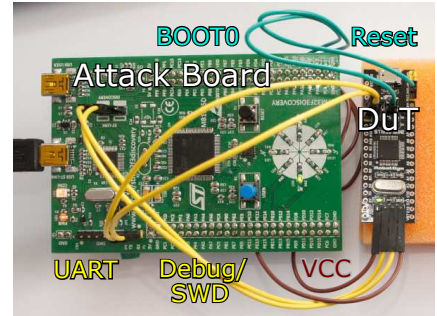

Figure 9: The setup for glitch generation, boot mode selection, and code extraction.

For the attack, we setup the DuT as shown in Figure 9. At first, we upload a two-stage exploit firmware to the SRAM with a debugger and shut down the debugger afterwards. Next, we configure SRAM-booting via the BOOT pins.

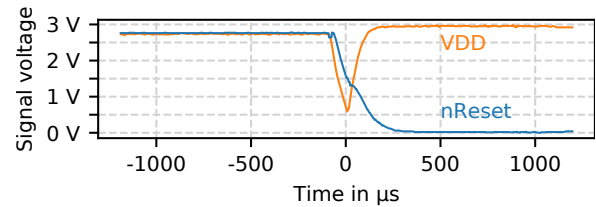

Figure 10: The VDD/supply glitch triggers a system reset.

However, the flash memory is still locked down since a debugger was connected. To release the lock down, we perform a very quick power cycle by cutting the power supply for a few hundreds of microseconds. During that glitch, we observe the microcontroller's reset line to detect when the system goes into reset. At this moment, our setup automatically restores power, as shown in Figure 10. After this very short power-cycle, access to the flash memory is re-enabled. But due to SRAM data remanence [18], such a short interruption in the power supply leaves the data in SRAM intact.

Next, we boot from SRAM and the stage one firmware configures the FPB. It is set to hijack the next device start up from flash memory by patching the reset vector fetch from the hard-coded address `0x00000004` [3]. We configure the FPB to replace this actual entry point in flash memory with the entry point of stage two of our shell code in SRAM.

Finally, we configure the chip to boot from flash memory and apply a reset. Since the FPB configuration is preserved, the patch configuration remains present. When the core tries to start up and fetches the entry address, this triggers the FPB which returns our alternative entry point. Thus, the control flow becomes immediately diverted to stage two in SRAM and a privileged shell is spawned. Flash memory remains accessible because the chip has – only seemingly – booted from the flash memory. The shell code has full system access and can extract the entire firmware in negligible time.

For the attack, we employ solely an STM32F303 evaluation board and some wires. The board provides a debug interface for the DuT, a UART interface for serial communication with the shell, and it controls the Reset, BOOT, and power supply signals via its GPIO pins. Two GPIOs are connected in parallel to provide sufficient current for the DuT. This experiment shows that low-cost tools are entirely sufficient to circumvent firmware protection in the STM32F103 and APM32F103.

## 8   Countermeasures

All the vulnerabilities emerge from the hardware design and implementation, thus, fixing these issues requires a new hardware revision. Most likely, no bullet-proof firmware-based workarounds exist for all those vulnerabilities, especially since their execution could be hampered by keeping the system in reset, if possible.

However, the device manufacturers are able to fix their devices in an updated *hardware* revision. Most of the vulnerabilities, especially concerning the debug interface and FPB, might require only minor modifications. For the invasive attacks on the dual-die GD32 devices, countermeasures are more demanding. The data could either be sufficiently encrypted or the manufacturer could switch from bonding wires to flip chip mounting of the flash chip. This prevents easy access to the inter-die connection, as tiny solder dots directly connect both dies at their area of contact.

To prevent vulnerabilities in the future, at least basic security reviews and tests are required during chip development. Some of the vulnerabilities were discovered by us within a few hours in a black-box scenario, thus, a person doing a white box test and having access to internal design documents should be able to discover most issues with ease. However, the underlying issue might less be on the technical side but more in the project management and requirement prioritization.

## 9   Conclusion and Outlook

In this paper, we analyzed drop-in replacements and the original STM32F103 microcontroller resulting in a total of six devices by four manufacturers. We showed via low-level analysis and optical inspection that the replacement devices are re-implementations of the STM32F103 and likely not based on illegitimately obtained design data. We focused on the flash memory security in each of those devices and came to alarming results especially because we limited ourselves to low-cost DiY approaches. When the replacement devices are compared to the STM32F103, some vulnerabilities appear to be fixed, but other, new vulnerabilities, affect those devices and expose the firmware nevertheless. Although there is no *magic exploit* affecting all devices at once, we still discovered shared vulnerabilities that affect multiple devices. Some design errors are observed across several chips, manufacturers, and even CPU architectures.

In total, the vulnerabilities allow extracting 100 % of the firmware from *every* tested device with comparably simple DiY methods. All the vulnerabilities are especially critical due to their easy exploitation which does not require a sophisticated lab. Hence, we identified a huge demand for security improvements in all of these microcontrollers, in the original and the replacements, to prevent firmware extraction and thereby an exposure of valuable data to adversaries and competitors. We strongly believe that the vast majority of those issues could have easily been identified in a security concept review or a penetration test before starting mass production.

However, these results could only be the tip of the iceberg, as we tested only the most popular series but security concepts and implementations are sometimes reused widely, also for other products. Additionally, while these vulnerabilities are obviously only design errors, we cannot exclude that systems might include even further, yet undiscovered, issues or even intentionally added and well-hidden backdoors in the worst case. In combination with the risk due to counterfeit STM32F103 devices, supply chain management becomes another vital factor for security. Thus, both, the trustworthiness of the manufacturers and the supply chain, play a major role in the resulting overall system security.

## 10   Coordinated Disclosure

We triggered a coordinated disclosure process with each manufacturer of the analyzed microcontrollers; i.e., Apex Microelectronics (APM), GigaDevice (GD), China Key Systems (CKS), and STMicroelectronics (ST). The details regarding the vulnerabilities of their microcontrollers were shared to the full extend more than 90 days prior to publication.

## 11   Supplementary Material

Supplementary material and code is available under https://science.obermaier-johannes.de/f103-analysis. It contains high-resolution chip die images, code for Cortex-M3 CPU revision and errata detection, logic traces during QSPI-memory communication (H1) and fault injection (H2), and code, as well as binaries, for the reproduction of all Proofs-of-Concepts D1-A, D1-B, D1-C, D2, H1, H2, H3.

## A Die Size Measurement

Table 3: Die sizes of each device for Flash (F) and Logic (L).

| Device | Die size in mm × mm | Die type |
|---|---|---|
| STM32F103(C8T6) | 3.3 × 3.3 | F + L |
| APM32F103(CBT6) | 3.0 × 2.6 | F + L |
| CKS32F103(C8T6) | 2.7 × 2.6 | F + L |
| GD32F103(C8T6) | 1.5 × 1.0 | F |
|  | 3.0 × 3.0 | L |
| GD32F130(C8T6) | 1.5 × 1.0 | F |
|  | 2.4 × 2.3 | L |
| GD32VF103(CBT6) | 1.5 × 0.8 | F |
|  | 3.1 × 3.2 | L |

## B Address Permutation Matrices

### B.1 GD32F103

$$\mathbf{P}_{\text{aPL}} = \begin{bmatrix} 0 & 0 & 0 & 0 & 0 & 0 & 0 & 1 \\ 0 & 0 & 0 & 0 & 0 & 0 & 1 & 0 \\ 0 & 0 & 0 & 0 & 0 & 1 & 0 & 0 \\ 0 & 0 & 0 & 0 & 1 & 0 & 0 & 0 \\ 0 & 0 & 0 & 1 & 0 & 0 & 0 & 0 \\ 0 & 0 & 1 & 0 & 0 & 0 & 0 & 0 \\ 0 & 1 & 0 & 0 & 0 & 0 & 0 & 0 \\ 1 & 0 & 0 & 0 & 0 & 0 & 0 & 0 \end{bmatrix}$$

### B.2 GD32F130

$$\mathbf{P}_{\text{aPL}} = \begin{bmatrix} 0 & 1 & 0 & 0 & 0 & 0 & 0 & 0 \\ 0 & 0 & 0 & 0 & 0 & 0 & 1 & 0 \\ 0 & 0 & 0 & 0 & 1 & 0 & 0 & 0 \\ 0 & 0 & 1 & 0 & 0 & 0 & 0 & 0 \\ 1 & 0 & 0 & 0 & 0 & 0 & 0 & 0 \\ 0 & 0 & 0 & 0 & 0 & 0 & 0 & 1 \\ 0 & 0 & 0 & 0 & 0 & 1 & 0 & 0 \\ 0 & 0 & 0 & 1 & 0 & 0 & 0 & 0 \end{bmatrix}$$

## C Bit Permutation Matrices

$$\mathbf{P}_\alpha = \begin{bmatrix} 0 & 0 & 0 & 0 & 1 & 0 & 0 & 0 \\ 1 & 0 & 0 & 0 & 0 & 0 & 0 & 0 \\ 0 & 0 & 0 & 0 & 0 & 0 & 0 & 1 \\ 0 & 0 & 0 & 0 & 0 & 0 & 1 & 0 \\ 0 & 0 & 1 & 0 & 0 & 0 & 0 & 0 \\ 0 & 0 & 0 & 1 & 0 & 0 & 0 & 0 \\ 0 & 0 & 0 & 0 & 0 & 1 & 0 & 0 \\ 0 & 1 & 0 & 0 & 0 & 0 & 0 & 0 \end{bmatrix}$$

$$\mathbf{P}_\beta = \begin{bmatrix} 1 & 0 & 0 & 0 & 0 & 0 & 0 & 0 \\ 0 & 0 & 0 & 0 & 0 & 0 & 0 & 1 \\ 0 & 0 & 0 & 0 & 0 & 0 & 1 & 0 \\ 0 & 0 & 1 & 0 & 0 & 0 & 0 & 0 \\ 0 & 0 & 0 & 1 & 0 & 0 & 0 & 0 \\ 0 & 0 & 0 & 0 & 0 & 1 & 0 & 0 \\ 0 & 1 & 0 & 0 & 0 & 0 & 0 & 0 \\ 0 & 0 & 0 & 0 & 1 & 0 & 0 & 0 \end{bmatrix}$$

$$\mathbf{P}_\gamma = \begin{bmatrix} 0 & 0 & 0 & 0 & 0 & 0 & 0 & 1 \\ 0 & 0 & 0 & 0 & 0 & 0 & 1 & 0 \\ 0 & 0 & 1 & 0 & 0 & 0 & 0 & 0 \\ 0 & 0 & 0 & 1 & 0 & 0 & 0 & 0 \\ 0 & 0 & 0 & 0 & 0 & 1 & 0 & 0 \\ 0 & 1 & 0 & 0 & 0 & 0 & 0 & 0 \\ 0 & 0 & 0 & 0 & 1 & 0 & 0 & 0 \\ 1 & 0 & 0 & 0 & 0 & 0 & 0 & 0 \end{bmatrix}$$

$$\mathbf{P}_\delta = \begin{bmatrix} 0 & 0 & 0 & 0 & 0 & 1 & 0 & 0 \\ 0 & 0 & 1 & 0 & 0 & 0 & 0 & 0 \\ 0 & 0 & 0 & 1 & 0 & 0 & 0 & 0 \\ 0 & 0 & 0 & 0 & 0 & 1 & 0 & 0 \\ 0 & 1 & 0 & 0 & 0 & 0 & 0 & 0 \\ 0 & 0 & 0 & 0 & 1 & 0 & 0 & 0 \\ 1 & 0 & 0 & 0 & 0 & 0 & 0 & 0 \\ 0 & 0 & 0 & 0 & 0 & 0 & 0 & 1 \end{bmatrix}$$

# D  Analyzed Devices

Table 4: Analyzed microcontrollers with their respective vulnerabilities and CVE numbers.

| Manufacturer | Device family | Device under test | Vulnerabilities | CVE |
|---|---|---|---|---|
| Apex Microelectronics | APM32F103 | APM32F103CBT6 [1] | **D1-B**: Extraction via Exceptions<br>**H3**: Glitch and FPB → Shellcode | CVE-2020-13463<br>CVE-2020-13471 |
| China Key Systems | CKS32F103 | CKS32F103C8T6 [6] | **D1-A**: Load Instruction Exploitation<br>**D2**: DMA Access Exploitation<br>**D1-B**: Extraction via Exceptions | CVE-2020-13464<br><br>CVE-2020-13467 |
| GigaDevice | GD32F103 | GD32F103C8T6 [10] | **D1-C**: Control Flow Redirection<br>**D2**: DMA Access Exploitation<br>**H1**: Invasive Data Eavesdropping | CVE-2020-13465<br>CVE-2020-13472<br>CVE-2020-13470 |
| GigaDevice | GD32F130 | GD32F130C8T6 [11] | **H1**: Invasive Data Eavesdropping<br>**H2**: Invasive RDP Manipulation | CVE-2020-13470<br>CVE-2020-13468 |
| GigaDevice | GD32VF103 | GD32VF103CBT6 [9] | **D1-A**: Load Instruction Exploitation | CVE-2020-13469 |
| STMicroelectronics | STM32F103 | STM32F103C8 [20] | **D1-B**: Extraction via Exceptions<br>**H3**: Glitch and FPB → Shellcode | CVE-2020-8004<br>CVE-2020-13466 |

**Important:** If a vulnerability is not listed for a specific device, one should not automatically assume its immunity. We observed that minor modifications to an attack approach can greatly influence the outcome. This table lists only vulnerabilities that were tested *and* successfully exploited. Debug interface attacks were tested for all devices. Invasive attacks were only tested extensively for selected devices. An adversary would not choose to mount an attack if there is an easy-to-exploit debug interface vulnerability, thus, the GD32VF103 was not tested against H1 and H2.